\pdfoutput=1
\documentclass[11pt]{article}
\usepackage{authblk}
\usepackage[final]{acl}

\usepackage{times}
\usepackage{latexsym}

\usepackage{colortbl}
\usepackage{comment}
\usepackage{adjustbox}
\usepackage{booktabs}
\usepackage{arydshln}
\usepackage{multirow}

\usepackage{amssymb}% http://ctan.org/pkg/amssymb
\usepackage{pifont}% http://ctan.org/pkg/pifont

\newcommand{\cmark}{\ding{51}}

\usepackage{graphicx}
\usepackage{amsmath}
\usepackage{amssymb}
\usepackage{booktabs}
\usepackage{algorithm}
\usepackage{listings}
\usepackage{placeins}
\usepackage[accsupp]{axessibility}
\usepackage{nccmath}

\NewDocumentCommand\emojismiley{}{
    \includegraphics[scale=0.045]{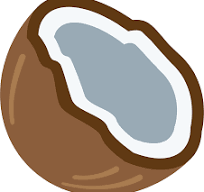}
}

\DeclareMathOperator*{\argmax}{arg\,max}

\usepackage{caption} 
\usepackage{subcaption}

\newcommand{\CC}[1]{\cellcolor{#1}}
\definecolor{ftcolor}{rgb}{1,1,1} % almond
\definecolor{baselinecolor}{rgb}{1,1,1}
\definecolor{contrcolor}{rgb}{1.0, 0.98, 0.8}
\definecolor{predcolor}{rgb}{0.74, 0.83, 0.9}
\definecolor{decorrcolor}{rgb}{0.98, 0.85, 0.87} % palepink
\definecolor{knowcolor}{rgb}{0.80, 0.94, 0.75}
\definecolor{offlinecolor}{rgb}{1,1,1}
\definecolor{firsttaskcolor}{rgb}{0.97, 0.97, 0.97}
\definecolor{azure(colorwheel)}{rgb}{0.0, 0.5, 1.0}
\definecolor{gray(x11gray)}{rgb}{0.75, 0.75, 0.75}
\definecolor{lightgray}{rgb}{0.90, 0.90, 0.90}
\definecolor{darkgray}{rgb}{0.66, 0.66, 0.66}

\definecolor{almondlow}{RGB}{252,239,219} 
\definecolor{almondmiddle}{RGB}{237,225,206} 
\definecolor{almondhigh}{RGB}{224,213,194} 
\definecolor{almondultra}{RGB}{214,204,186}

\definecolor{pastelred}{RGB}{232, 131, 131}
\definecolor{pastelviolet}{rgb}{0.8, 0.7, 0.79}
\definecolor{champagne}{rgb}{0.97, 0.91, 0.81}
\usepackage[T1]{fontenc}
\usepackage[utf8]{inputenc}

\usepackage{microtype}

\usepackage{inconsolata}

\DeclareSymbolFont{extraup}{U}{zavm}{m}{n}
\DeclareMathSymbol{\varheart}{\mathalpha}{extraup}{86}
\DeclareMathSymbol{\vardiamond}{\mathalpha}{extraup}{87}

\title{Continual Contrastive Spoken Language Understanding}

\author[$\vardiamond$]{\textbf{Umberto Cappellazzo}\textbf{\dag}}
\author[$\heartsuit$]{\textbf{Enrico Fini}}
\author[$\spadesuit$]{\textbf{Muqiao Yang}}
\author[$\clubsuit$]{\textbf{Daniele Falavigna}}
\author[$\clubsuit$]{\textbf{Alessio Brutti}\thanks{This paper has received funding from the European Union’s Horizon research and innovation programme under grant agreement No 101135798, project Meetween (My Personal AI Mediator for Virtual MEETtings BetWEEN People)}}
\author[$\spadesuit$]{\textbf{Bhiksha Raj}}

\affil[$\vardiamond$]{University of Trento}
\affil[$\heartsuit$]{Independent Researcher}
\affil[$\spadesuit$]{Carnegie Mellon University}
\affil[$\clubsuit$]{Fondazione Bruno Kessler}

\affil[ ]{\textbf{\dag}~Corresponding author: \textit {umberto.cappellazzo@unitn.it}}

\begin{document}
\maketitle
\begin{abstract}
Recently, neural networks have shown impressive progress across diverse fields, with speech processing being no exception. However, recent breakthroughs in this area require extensive offline training using large datasets and tremendous computing resources. Unfortunately, these models struggle to retain their previously acquired knowledge when learning new tasks continually. In this paper, we investigate the problem of learning sequence-to-sequence models for spoken language understanding in a class-incremental learning (CIL) setting and we propose COCONUT \emojismiley, a CIL method that relies on the combination of experience replay and contrastive learning. Through a modified version of the standard supervised contrastive loss, COCONUT preserves the learned representations by pulling closer samples from the same class and pushing away the others. Moreover, we leverage a multimodal contrastive loss that helps the model learn more discriminative representations of the new data by aligning audio and text features. We also investigate different contrastive designs to combine the strengths of the contrastive loss with teacher-student architectures used for distillation. Experiments on two established SLU datasets reveal the effectiveness of our proposed approach and significant improvements over the baselines. We also show that COCONUT can be combined with methods that operate on the decoder side, resulting in further metrics improvements.
\end{abstract}

\section{Introduction}

With the rapid progress of intelligent
voice-enabled personal assistants, the significance of Spoken Language Understanding (SLU) has gained substantial recognition in recent years \citep{arora2022espnet, qin2021survey}. Conventional SLU models deploy a cascaded pipeline of an automatic speech recognition (ASR) system followed by a natural language understanding (NLU) module \citep{mesnil2014using, Horlock2003DiscriminativeMF}. ASR maps the input speech into text representations, and NLU extracts the target intent labels from the intermediate text. Even though these approaches can leverage a vast abundance of ASR and NLU data, 
they suffer from ASR error propagation. Conversely, end-to-end (E2E) SLU \citep{agrawal2022tie, lugosch2019speech, saxon2021end} has received more attention in recent research because it uses a single trainable model to map the speech audio directly to the intent labels, bypassing %the need to explicitly generate a 
the text transcript %. This approach leads to 
and reducing latency and error propagation. 

%\textcolor{red}{*}E2E methods also allow the integration of pre-trained ASR and language models (LMs) to overcome the dearth of labeled data for SLU \citep{arora2023integrating, end-to-end2}\textcolor{red}{*}.

The assumption that the data distribution the model will face after deployment aligns with what it encountered during the training phase is brittle and unrealistic. In fact, real-world scenarios entail evolving streams of data where novel categories (e.g., new vocabulary or intents) emerge sequentially, known as continual learning (CL). Unfortunately, while neural networks thrive in a stationary environment, the situation is reversed in CL, resulting in the ``catastrophic forgetting'' (CF) of the existing knowledge in favor of fresh new information \citep{mccloskey1989catastrophic}. Although the majority of CL works have focused on computer vision tasks like image classification \citep{buzzega2020dark, wang2022learning} and semantic segmentation \citep{maracani2021recall, yang2022uncertainty}, a few works have recently turned their attention towards text \citep{wang2023effective, ke2023continual} and speech \citep{cappellazzoinvestigation, diwan2023continual}, as well as vision-language \citep{ni2023continual,zhu2023ctp} and vision-audio \citep{mo2023class, pian2023audio}.

While most SLU works consider offline settings, a thorough study of SLU under a class-incremental learning (CIL) setup still lacks. In CIL, one single model is adapted to a sequence of different tasks as incremental labels emerge sequentially. Recently, \citet{cappellazzo2023sequence} studied the problem of CIL in ASR-SLU, where SLU is carried out in a sequence-to-sequence (seq2seq) fashion, thus computing the intent labels in an auto-regressive way together with the ASR transcriptions. By doing this, the model comprises three blocks: text and audio encoders, and an ASR decoder. %While \citet{cappellazzo2023sequence} 
While in that work the knowledge distillation (KD) principle applied to the ASR decoder is used, in this paper, we exploit the multi-modal audio-text setting and propose \textbf{COCONUT} \emojismiley: \textbf{CO}ntinual \textbf{C}ontrastive sp\textbf{O}ken la\textbf{N}guage \textbf{U}nders\textbf{T}anding. COCONUT combines experience replay (ER) and contrastive learning principles. Whereas ER is a well-established approach in CL, whereby a bunch of old training samples are collected into a dedicated rehearsal memory buffer and interleaved with the data from the new task \cite{rolnick2019experience, bang2021rainbow}, only recently has contrastive learning been harnessed to learn representations continually. Both supervised \citep{cha2021co2l, yang2022uncertainty} and self-supervised \citep{fini2022self, wang2022learning} contrastive learning have proven useful to lessen the CF issue. Specifically, COCONUT relies on two contrastive learning-based losses that operate on a shared embedding space where the audio and text features are projected. 

The first loss coined \textit{Negative-Student Positive-Teacher} (NSPT), is a modified version of the supervised contrastive learning loss that aims to consolidate what the model has learned in the previous tasks. It also exploits KD \citep{hinton2015distilling, li2017learning} to guide the current model (student) to produce representations that resemble the ones obtained with the model from the previous tasks (teacher). \textit{For this reason, this loss is computed only on the rehearsal data (i.e., the anchors)}. A key difference between our loss and the standard contrastive one is that the positive samples are computed using the teacher (the positives only come from the rehearsal data), whereas the negatives are computed with the student. In this way, we avoid stale and scattered representations for the new data.

\begin{figure*}[t]
    \centering
    \includegraphics[width=0.9\textwidth]{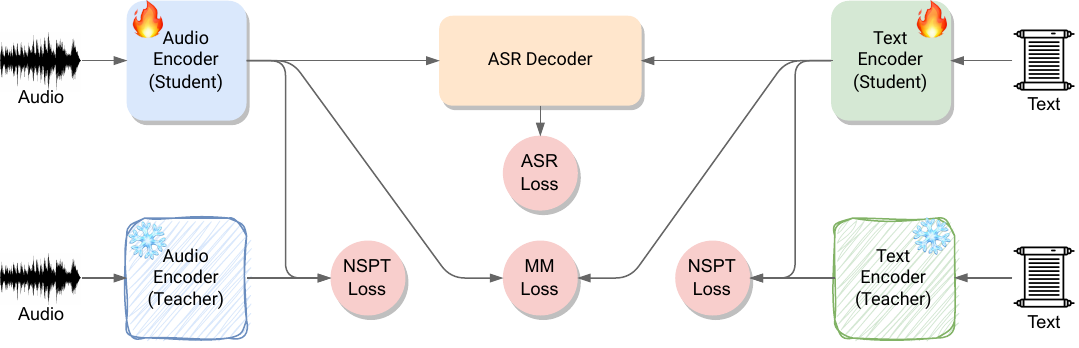}
    \caption{Overview of COCONUT \emojismiley. It uses two contrastive learning-based losses. The NSPT (negative-student positive-teacher) loss is a supervised contrastive distillation loss that preserves the feature representations of the \textit{past} classes for both audio and text samples. The positive and negative samples are computed with the teacher and student model, respectively. The MM (multi-modal) loss aims to align audio and text representations belonging to the same \textit{new} class. COCONUT produces features that are more transferable and resilient to catastrophic forgetting.}
    \label{fig:method}
\end{figure*}

The second loss is inspired by the recent progress in multi-modal representation learning. Considering that for audio-text paired data, audio and text represent the same information but in different ways, it has been shown that aligning their representations results in better performance for various speech-related problems \citep{zhu2022cross, ye2022cross, manco2022contrastive}. Therefore, we propose a multi-modal (MM) supervised contrastive loss that, \textit{exclusively applied to the current task's data}, brings audio and text representations belonging to the same class into closer proximity in the shared feature space, resulting in features that are more transferable and resilient to CF. An overview of COCONUT is illustrated in Figure~\ref{fig:method}.

In summary, our contributions are three-fold: \ding{182} we introduce COCONUT \emojismiley, a CL method that makes use of two supervised contrastive learning objectives to mitigate CF for seq2seq SLU models. In particular, through our proposed NSPT loss we provide a detailed study of which models (student/teacher) should be used at the numerator/denominator (positives/negatives) of the contrastive loss tailored for class-incremental learning. \ding{183} We conduct extensive experiments on two popular SLU benchmarks demonstrating that COCONUT achieves consistent improvements over the baselines. We also show that it can be combined with KD applied to the ASR decoder, leading to further improvements. Finally, \ding{184} we ablate the contribution of each loss and its components, showcasing their pivotal role in COCONUT. %achieving the best results. 

\section{Problem Formulation}

\subsection{ASR-SLU Multi-task Learning} 
SLU is considered a more difficult task than ASR and NLU since it involves both acoustic and semantic interpretation \citep{tur2011spoken}. For this reason,  
it is common practice to include an additional ASR objective such that the SLU labels (in our case the intent labels) and the transcript are generated in an auto-regressive fashion, resulting in a multi-task learning setting \citep{arora2022espnet, peng2023study}. By doing this, the text transcript input to the model includes a class intent token that is specific to the actual task.

Let $\theta$ be the parameters of a seq2seq ASR model comprising an audio encoder, a text encoder (i.e., embedding layer), and an ASR decoder. Let $\textbf{x} = [x_0,\dots,x_{U-1}]$ be an audio input sequence of length $U$, and $\textbf{y} = [y_{cls},y_{sep}, y_0,\dots,y_{J-3}]$ be the ``extended'' input transcript of length $J$, where with the term ``extended'' we refer to the original transcript $ [y_0,\dots,y_{J-3}]$ augmented with the intent class token $y_{cls}$ and a special separation token $y_{sep}$. The goal of the ASR model is to find the most likely extended transcript given the input sequence $\textbf{x}$:
\begin{equation}
   \hat{\textbf{y}} = \argmax_{\textbf{y} \in \mathcal{Y}^*} p(\textbf{y}|\textbf{x};\theta),
\end{equation}
where $\mathcal{Y}^*$ is the set of all token sequences. The predicted intent is obtained extracting $y_{cls}$ from $\hat{\textbf{y}}$.
\subsection{Class-Incremental Learning}
For our experiments, we consider a CIL setting where we adapt a single model to learn sequentially $N$ tasks corresponding to non-overlapping subsets of classes (in our case \textit{intents}). Put formally, the training dataset is divided into $N$ distinct tasks, $\mathcal{D}=\{\mathcal{D}_0,\ldots,\mathcal{D}_{N-1}$\}, based on the intent token $y_{cls}$, so that one intent is included in one and only one task. The dataset $\mathcal{D}_n$ of task $n$ comprises audio signals $\mathcal{X}_n$ with associated transcriptions $\mathcal{Y}_n$, i.e. $\mathcal{D}_n=(\mathcal{X}_n,\mathcal{Y}_n)$. The CIL setting is challenging in that the model must be able to distinguish all classes until task $n$, thus at inference time the task labels are not available (unlike in task-incremental learning) \citep{hsu2018re}.

\section{Proposed Approach}

\subsection{Standard Rehearsal-based Approach}

We assume the availability of a rehearsal buffer, $\mathcal{M}$, in which we can store a few samples for each class encountered in the previous tasks. During the training phase of task $n$, $\mathcal{D}_n$, we refer to $\mathcal{B}$ as a mini-batch of samples $(\textbf{x},\textbf{y})$, some of which come from the current task and others from the rehearsal memory. 
%The rehearsal memory is gradually filled up until we reach the last task, so we do not need to employ any replacement strategy to make room for the classes of the new task.
To increase the variance of the audio data, we apply SpecAug \citep{park2019specaugment} to the audio waveform $\textbf{x}$ %as a data augmentation transformation 
(see \ref{sec:specaug} for more details). We do not implement any augmentation technique for the transcript $\textbf{y}$.
%Then, 
We encode each modality separately through a dedicated feature encoder. An audio encoder
maps each audio input into a feature vector $\textbf{h}_{\operatorname{A}} \in \mathbb{R}^{U \times d_{\operatorname{A}}}$, where $d_{\operatorname{A}}$ is the audio hidden size. Similarly, a text encoder converts each text input into a feature vector $\textbf{h}_{\operatorname{T}} \in \mathbb{R}^{J \times d_{\operatorname{T}}}$, where $d_{\operatorname{T}}$ is the text hidden size. At this point, if no specific CL losses are used, the ASR decoder generates the output sequence in an auto-regressive fashion, cross-attending on the audio encoder's representations $\textbf{h}_{\operatorname{A}}$. Thus, at task $n$, we minimize the conventional cross-entropy loss over the current mini-batch $\mathcal{B}$: 
\begin{equation}
    \mathcal{L}_{\text{ASR}} =- \frac{1}{|\mathcal{B}|}\sum_{(\textbf{x},\textbf{y})\in \mathcal{B}}\log(p(\textbf{y}|\textbf{x};\theta)).
\label{eq:ce}
\end{equation}
\subsection{COCONUT \emojismiley}
\label{sec:coconut}
\textbf{Preliminaries}. We introduce here some notations for our proposed approach. % COCONUT \emojismiley. 
Since we work with audio and text sequences, we need to aggregate the features we obtain with the encoders before computing the contrastive loss. For the audio component $\textbf{h}_{\operatorname{A}}$ we apply a mean operation over its sequence length, whereas for text we only select the feature related to the intent token. Then, as is common practice in contrastive learning \citep{radford2021learning, chen2020simple}, the resulting embeddings go through two separate linear projection layers that map them into a shared embedding space. At inference time, the projection layers are discarded. Therefore, we get the projected embeddings $\textbf{a}$ and $\textbf{t}$ in the following way:
\begin{equation}
    \textbf{a} = g_{\operatorname{A}}(avg(\textbf{h}_{\operatorname{A}})), \quad \textbf{t} = g_{\operatorname{T}}(cls(\textbf{h}_{\operatorname{T}})),
\end{equation}
where $cls(\cdot)$ is a function that extracts the feature associated with the class token, $g_{\operatorname{A}}(\cdot)$ and $g_{\operatorname{T}}(\cdot)$ are the projection layers, $\textbf{a} \in \mathbb{R}^{d_{\operatorname{S}}}$ and  $\textbf{t} \in \mathbb{R}^{d_{\operatorname{S}}}$, where $d_{\operatorname{S}}$ is the dimension of the shared space. 

Furthermore, we introduce some notations for the indices of samples coming from the current mini-batch $\mathcal{B}$. Let $\mathcal{I}_c$ and $\mathcal{I}_r$ represent the set of indices of the \textit{new task} samples and the indices of the samples from the rehearsal memory (\textit{old task} samples) in $\mathcal{B}$, respectively. Also, let $\mathcal{I} = \mathcal{I}_c \cup \mathcal{I}_r$, and we define $\mathcal{P}(k)$ as the set of indices of positive samples (i.e., samples with the same intent token). 

\begin{figure*}[t]
    \centering
    \includegraphics[width=0.9\textwidth]{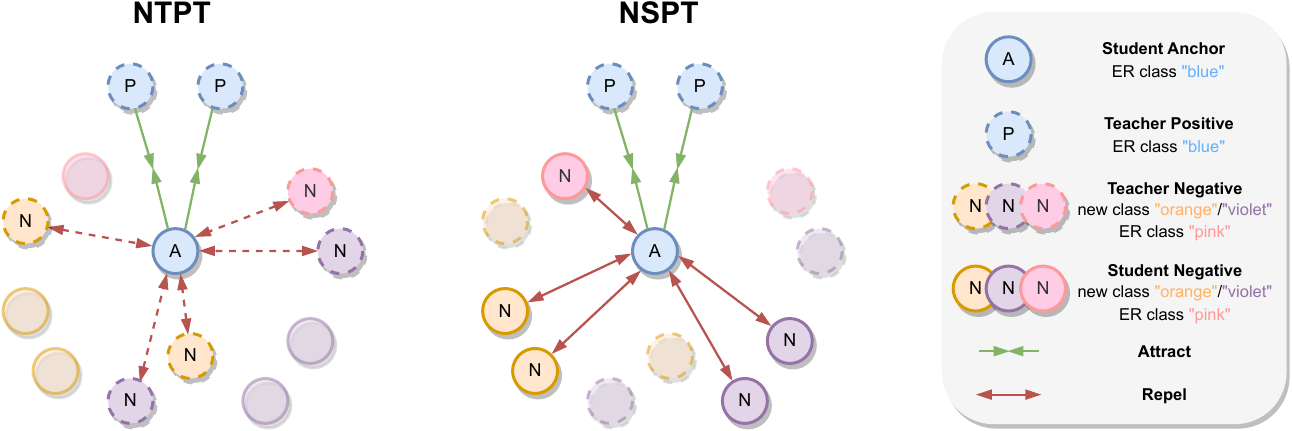}
    \caption{Illustration of the NTPT loss and our proposed NSPT loss. Given an anchor sample from the current mini-batch, the NTPT loss computes the negatives and positives using the teacher model (dashed circles). Instead, the NSPT loss computes the positives with the teacher while the negatives are computed with the student model (solid circles). If the features obtained with the teacher are scattered and static (the teacher is frozen), those obtained with the student are more clustered and can be learned during the current task. Best viewed in color.}
    \label{fig:NSPT}
\end{figure*}

The objective of a standard supervised contrastive loss (SCL) \citep{khosla2020supervised} is to push the representations of samples with different classes (negative pairs) farther apart while clustering representation of samples with the same class (positive pairs) closely together. Suppose that we get from the projection layers a generic representation $\textbf{z}_i^D$ for the $i$-th element in the batch, where $\textbf{z}=\{\textbf{a}, \textbf{t}\}$ and the superscript $D$ denotes whether the representation is computed with the teacher or student model. A generic formulation of the SCL loss takes the following form:
\begin{equation}
    \medmath{\mathcal{L}_{\text{SCL}} = \sum_{k \in \mathcal{I}} \frac{-1}{\vert \mathcal{P}(k) \vert} \sum_{p \in \mathcal{P}(k)} 
    \log \frac{\exp(\textbf{z}^{D}_k 
    \cdot \textbf{z}^{D}_p/\tau)}{\sum_{i \in \mathcal{I} }\exp(\textbf{z}^{D}_k \cdot \textbf{z}^{D}_i/\tau)}},
\end{equation}
$\tau \in \mathbb{R}^+$ is a fixed temperature scaling parameter.

\textbf{Supervised Contrastive Distillation Loss (NSPT)}. 
This loss combines the benefits of KD with those of contrastive learning \citep{tian2019contrastive, sun2020contrastive}. We recall that KD-based methods are very popular in CL and they exploit this paradigm to penalize changes to the model’s intermediate or final outputs by fostering the pass of the knowledge accrued in the teacher model onto the student \cite{rebuffi2017icarl, douillard2020podnet, cappellazzoinvestigation}. Commonly, we denote with \textit{teacher} the model trained in the previous task, and with \textit{student} that trained in the current task. Therefore, since the teacher conveys information about the previous classes, we would like to use it as a guide for the student through a KD objective. In this way, the loss encourages the student to produce audio and text embeddings consistent with those obtained by the teacher. For this reason, only the rehearsal samples are involved in this process as the teacher had no chance to see the current data. Additionally, we want to pull closer embeddings sharing the same intent class (i.e. the positives), while we push away the others (i.e. the negatives, whose class is different). This is obtained via a modified version of the standard supervised contrastive loss tailored for our setting. In fact, a standard one would use the teacher to compute both the positives and the negatives \citep{khosla2020supervised}. However, since the teacher is frozen and it is pointless to compute the representations of the samples from the current task using the teacher, we propose to use the student for computing the representations of the negatives. A small fraction of negatives come from the rehearsal buffer, and we also compute them using the student. We show in section \ref{sec:ablation} that using the teacher deteriorates the performance. Therefore, our contrastive distillation loss computes the embeddings of the anchor and its corresponding negatives using the student, while the positives come from the teacher (we call this loss \textit{Negative-Student Positive-Teacher}, NSPT). On the contrary, for the standard contrastive loss both the positives and negatives are computed with the teacher (we call it  \textit{Negative-Teacher Positive-Teacher}, NTPT). Figure~\ref{fig:NSPT} illustrates visually how the NTPT and NSPT work in the shared embedding space. The NSPT loss is computed for both audio and text embeddings, leading to two components, one for each modality, as follows:
\begin{equation}
    \medmath{
    \begin{split}
    \mathcal{L}_{\text{NSPT}} =   \sum_{k \in \mathcal{I}_{r}} \frac{-1}{\vert \mathcal{P}(k) \vert} & \sum_{p \in \mathcal{P}(k)} \Biggl[ \; \underbrace{\log \frac{\exp(\textbf{a}^{n}_k \cdot \textbf{a}^{n-1}_p/\tau)}{\sum_{i \in \mathcal{I} }\exp(\textbf{a}^{n}_k \cdot \textbf{a}^{n}_i/\tau)}}_{\mathcal{L}_{\text{A}}} + \\ & \underbrace{\log \frac{\exp(\textbf{t}^{n}_k \cdot \textbf{t}^{n-1}_p/\tau)}{\sum_{i \in \mathcal{I} }\exp(\textbf{t}^{n}_k \cdot \textbf{t}^{n}_i/\tau)}}_{\mathcal{L}_{\text{T}}} 
\; \Biggl],
\end{split}}
\label{eq:nspt}
\end{equation}
where $n$ and $n-1$ denote whether the representation is obtained with the student or teacher, and 
$\mathcal{L}_{\text{A}}$ and $\mathcal{L}_{\text{T}}$ represent the audio and text contributions, respectively. 
We empirically validate that the intuition of the NSPT loss is beneficial in section~\ref{sec:ablation}.

\textbf{Supervised Multi-Modal Contrastive Loss}. 
This loss is introduced for two reasons. First of all, since during the first task (no CL) the NSPT loss is not computed (i.e., we do not have a teacher yet), this means that the projector layers of the model are not trained. This would be a problem from the second task onwards in that the student would distill the knowledge from the teacher with randomly initialized projectors. Second, we want to exploit the multi-modal nature of our SLU CIL setting. Consequently, we introduce a multi-modal (MM) loss that aims to align audio and text representations belonging to the same new class, and thus training the projectors of the model from the very beginning. This alignment is achieved via a supervised multi-modal (i.e., audio-text) contrastive learning objective where feature representations of samples sharing the same intent token are attracted while the others are pushed away. Similar to \citep{kwon2022masked}, we use the [CLS] text token ($y_{cls}$) for performing the multi-modal alignment. Furthermore, following \citep{cha2021co2l}, we always treat the rehearsal samples as negatives, preventing them from being anchors during the learning process. This design choice is buttressed by two motivations: \textbf{1)} rehearsal data have been learned by the previous model already and are preserved via the NSPT loss, and \textbf{2)} we encourage the model to produce clusters for the new data that are separated from those of the rehearsal data. The MM loss is defined as: 
 \begin{equation}
    \medmath{
    \begin{split}
    \mathcal{L}_{\text{MM}} = \sum_{k \in \mathcal{I}_c} \frac{-1}{\vert \mathcal{P}(k) \vert} & \sum_{p \in \mathcal{P}(k)} \Biggl[ \log \frac{\exp(\textbf{a}_k^n \cdot \textbf{t}_p^n/\tau)}{\sum_{i \in \mathcal{I} }\exp(\textbf{a}_k^n \cdot \textbf{t}_i^n/\tau)} + \\ & \log \frac{\exp(\textbf{t}_k^n \cdot \textbf{a}_p^n/\tau)}{\sum_{i \in \mathcal{I}}\exp(\textbf{t}_k^n \cdot \textbf{a}_i^n/\tau)}
    \Biggr].
    \end{split}}
\label{eq:mm}
\end{equation}
The first term of the internal loss is the audio-to-text component, whereas the second is the text-to-audio component \citep{zhang2022contrastive}. The presence of both directions ($A \rightarrow T$ and $T \rightarrow A$) makes the MM loss symmetric.
All in all, COCONUT minimizes the following loss: 
\begin{equation}
\mathcal{L} = \mathcal{L}_{\text{ASR}} + \lambda_\text{MM} \mathcal{L}_{\text{MM}} + \lambda_\text{NSPT} \mathcal{L}_{\text{NSPT}},
\end{equation}
where lambdas are loss-specific weights. Note that during the first task $\mathcal{L}_{\text{NSPT}}$ is not computed.

\section{Experiments}

\subsection{Experimental Setup and Implementation Details}
\textbf{Datasets and CIL setting}. We evaluate COCONUT on two SLU datasets: the Fluent Speech Commands (FSC) \citep{lugosch2019speech} and the Spoken Language Understanding Resource Package (SLURP) \citep{bastianelli2020slurp}. FSC includes 30,043 English utterances, recorded at 16 kHz, resulting in 31 intent classes in total. The SLURP dataset comprises around 56 hours of audio of people interacting with a home assistant (\textit{slurp\_real}), with the addition of 43.5 hours of synthetic data (\textit{slurp\_synth}). It is considered the most challenging SLU dataset due to its lexical complexity. Each utterance is annotated with 3 semantics: scenario, action, and entity. The pair (scenario, action) defines an intent. Overall, there are 18 scenarios and 69 intents. For our experiments, we only perform intent classification.  Following \citep{cappellazzo2023sequence}, we use the scenario labels as splitting criterion to define the CIL setting (we refer to \ref{sec:CIL} for more details on this). We experiment on two configurations: 1) the datasets are partitioned into 3 tasks, each task comprising 6 scenarios for SLURP (denoted as SLURP-3), and 10 intents for FSC (FSC-3); 2) a more challenging configuration with 6 tasks, each task including 3 scenarios for SLURP (SLURP-6), and 5 intents for FSC (FSC-6).

\begin{table*}[t]
\centering
    \caption{Results 
    in terms of Average Accuracy ($\uparrow$), Last Accuracy ($\uparrow$), and Average WER ($\downarrow$) for different strategies on FSC and SLURP datasets. All CL methods exploit a buffer whose size is $1$\% of the training dataset. \textbf{Bold} and \underline{underscore} numbers denote the best and second best method for a specific setting and metric, respectively.  We show in the last row that COCONUT and S-KD can be used together, leading to the best results. For simplicity, the values of the last row are not in bold even though attain the best results.}
    \begin{adjustbox}{width=\textwidth,center}
    \begin{tabular}{lcccccccccccc}
    \toprule
    
      \bf{Setting} $\rightarrow$ &  \multicolumn{3}{c}{\CC{pastelviolet}\textbf{FSC-3}} &   \multicolumn{3}{c}{\CC{pastelviolet}\textbf{FSC-6}} & \multicolumn{3}{c}{\CC{champagne}\textbf{SLURP-3}} &   \multicolumn{3}{c}{\CC{champagne}\textbf{SLURP-6}} \\
       & \multicolumn{3}{c}{---------------------------} & \multicolumn{3}{c}{---------------------------} & \multicolumn{3}{c}{---------------------------} & \multicolumn{3}{c}{---------------------------}\\

     \textbf{Metric} $\rightarrow$ & Avg & Last & Avg  & Avg & Last & Avg & Avg & Last & Avg  & Avg & Last & Avg\\ 
    \textbf{Method} $\downarrow$ & Acc &Acc & WER &Acc &Acc &WER &Acc &Acc & WER &Acc &Acc &WER\\ 
    
    \midrule
    
    \textcolor{darkgray}{Offline} & \textcolor{darkgray}{99.28} & \textcolor{darkgray}{-} &\textcolor{darkgray}{0.48}  &\textcolor{darkgray}{99.28} & \textcolor{darkgray}{-} &\textcolor{darkgray}{0.48} &\textcolor{darkgray}{84.41} & \textcolor{darkgray}{-} &\textcolor{darkgray}{17.65}  &\textcolor{darkgray}{84.41} & \textcolor{darkgray}{-} &\textcolor{darkgray}{17.65} \\ 
    Fine-tuning &49.13 &17.61 &36.37 &29.92 &7.59 &54.66 &46.65 &18.42 &28.32 &31.90 &10.57 &34.79\\ \addlinespace[2pt]
    \hline \addlinespace[2pt]
    %\textcolor{darkgray}{ER} &\textcolor{darkgray}{2\% / rand}&\textcolor{darkgray}{88.02} &\textcolor{darkgray}{84.20} &\textcolor{darkgray}{9.19}  &\textcolor{darkgray}{81.19} &\textcolor{darkgray}{79.71} &\textcolor{darkgray}{13.75} &\textcolor{darkgray}{75.62} &\textcolor{darkgray}{68.68} &\textcolor{darkgray}{19.55}  &\textcolor{darkgray}{72.75} &\textcolor{darkgray}{68.49} &\textcolor{darkgray}{22.98} \\
    ER rand &79.17 &69.81 &15.87 &68.61 &63.71 &24.04 &71.44 &61.88 &21.25 &66.57 &58.22 &24.50\\ 
    ER iCaRL &82.04 &74.00 &13.45 &69.76 &64.12 &23.22 &71.94 &63.22 &\underline{21.06} &68.08 &62.29 &26.05\\ \addlinespace[2pt]
    \hline \addlinespace[2pt]
    T-KD &82.11 &75.43 &12.95  &69.08 &64.73 &23.82 &72.44 &62.43 &21.19 &66.95 &60.47 &\textbf{24.26}\\ \addlinespace[2pt]
    A-KD &\underline{84.79} &\underline{78.12} &\underline{11.54} &73.54 &67.05 &\underline{20.36} &72.10 &63.84 &\textbf{20.67} &68.52 &62.51 &\underline{24.29}\\\addlinespace[2pt]
    S-KD &84.29 &75.31 &12.39 &\underline{73.65} &\underline{67.71} &21.27  &\textbf{74.28} &\textbf{65.95} &21.26 &\underline{69.91} &\underline{63.22} &\textbf{24.26}\\  \addlinespace[2pt]
    \CC{predcolor}\textbf{COCONUT} & \CC{predcolor}\textbf{86.39} &\CC{predcolor}\textbf{80.21} &\CC{predcolor}\textbf{11.08} &\CC{predcolor}\textbf{77.09} &\CC{predcolor}\textbf{73.80} &\CC{predcolor}\textbf{19.05} &\CC{predcolor}\underline{72.75} &\CC{predcolor}\underline{64.62} &\CC{predcolor}21.25 &\CC{predcolor}\textbf{70.17} &\CC{predcolor}\textbf{63.66} &\CC{predcolor}\underline{24.29}\\  \addlinespace[2pt]

    \hdashline  \addlinespace[2pt]
    \textcolor{predcolor}{\textbf{COCONUT+S-KD}} & \textcolor{predcolor}{87.64} &\textcolor{predcolor}{80.45} &\textcolor{predcolor}{10.49} &\textcolor{predcolor}{77.57} &\textcolor{predcolor}{74.01} &\textcolor{predcolor}{18.47} &\textcolor{predcolor}{75.58} &\textcolor{predcolor}{67.39} &\textcolor{predcolor}{20.61} &\textcolor{predcolor}{71.91} &\textcolor{predcolor}{65.41} &\textcolor{predcolor}{24.16} \\
    
    \bottomrule
    %\Xhline{2\arrayrulewidth}
    \end{tabular}
    \end{adjustbox}
    \label{tab:result_SLURPFSC}
\end{table*}

\begin{figure*}
     \centering
     \begin{subfigure}[b]{0.45\textwidth}
         \centering
         \includegraphics[width=\textwidth]{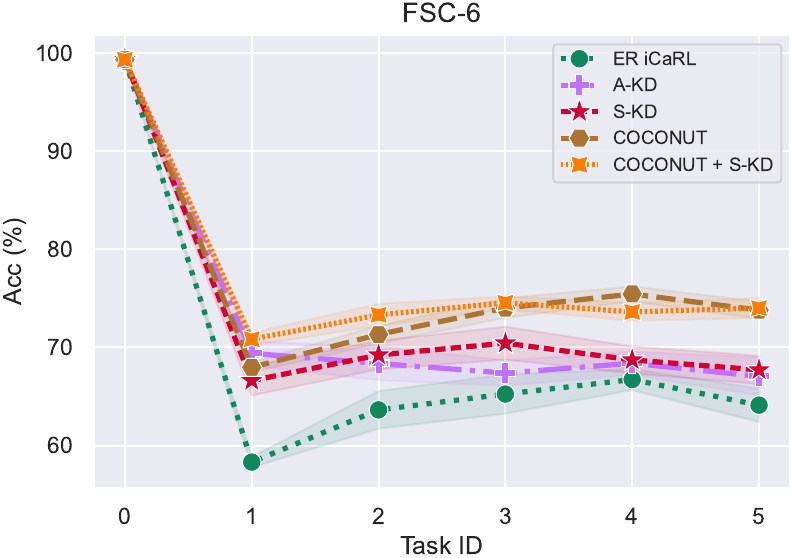}
         %\caption{$y=x$}
         %\label{fig:y equals x}
     \end{subfigure}
     %\hfill
     \begin{subfigure}[b]{0.45\textwidth}
         \centering
         \includegraphics[width=\textwidth]{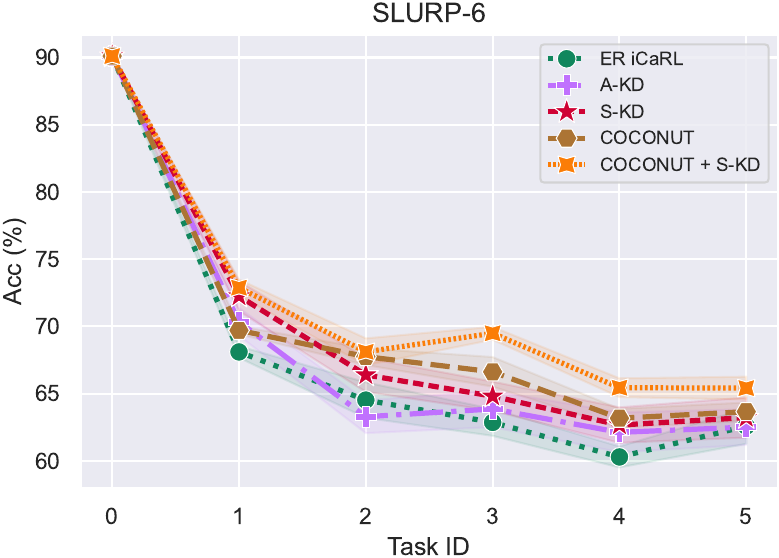}
         %\caption{$y=3\sin x$}
         %\label{fig:three sin x}
     \end{subfigure}
     %\begin{subfigure}[b]{0.32\textwidth}
     %    \centering
     %    \includegraphics[width=\textwidth]{figures_plots/fsc6_avgacc_varyingmem.pdf}
     %\end{subfigure}
    
    \caption{\textit{Left}: the trend of the intent accuracy on the observed tasks for the FSC-6 setting. \textit{Right}: the trend of the intent accuracy on the observed tasks for SLURP-6.}
    \label{fig:acc}
\end{figure*}

\textbf{Implementation Details}. For both datasets, the text encoder is a standard text embedding layer with size $768$. For the audio encoder, we use a Wav2vec 2.0 base model \citep{baevski2020wav2vec} pre-trained and fine-tuned on 960 hours of Librispeech for SLURP ($\sim 94.3$M parameters), while we use DistilHuBERT base \citep{chang2022distilhubert} for FSC ($\sim 23.5$M parameters). Both encoders have hidden sizes of $768$. Since FSC is a less challenging dataset than SLURP, we found that a smaller pre-trained encoder is sufficient to achieve state-of-the-art results. Moreover, experimenting with diverse architectures helps evaluate the generalizability of our proposed method. As in \citep{radford2021learning}, we employ linear projection layers to map from each encoder’s representation to the audio-text embedding space, whose dimension is $512$. The ASR decoder is transformer-based with $6$ layers, hidden size equal to $768$, $8$ attention heads, and the dimension of the feedforward layers is $2048$. We set the temperature $\tau$ to $0.1$ for both NSPT and MM loss (please refer to \ref{sec:temperature} for a detailed analysis).

For the tokenization we apply Byte-Pair Encoding (BPE) \citep{sennrich2016neural} for SLURP, with a vocabulary size of $1000$ and BPE dropout equal to $0.1$, whereas for FSC, given the limited number of unique words, we use word tokenization, resulting in 139 tokens. BPE automatically assigns to each intent a dedicated token, whereas for FSC we manually add the intent tokens. We refer the reader to \ref{sec:hyperparams} for an exhaustive description of the hyperparameters. Regarding the weight coefficients, we set $\lambda_{\text{MM}}$ to $0.1$, and similarly to \citep{douillard2022dytox, wu2019large} we set $\lambda_{\text{NSPT}}$ to $\frac{L_p}{L_p + L_n}$, where $L_p$ and $L_n$ count the number of past and new classes.

\begin{figure*}
     \centering
     \begin{subfigure}[b]{0.45\textwidth}
         \centering
         \includegraphics[width=\textwidth]{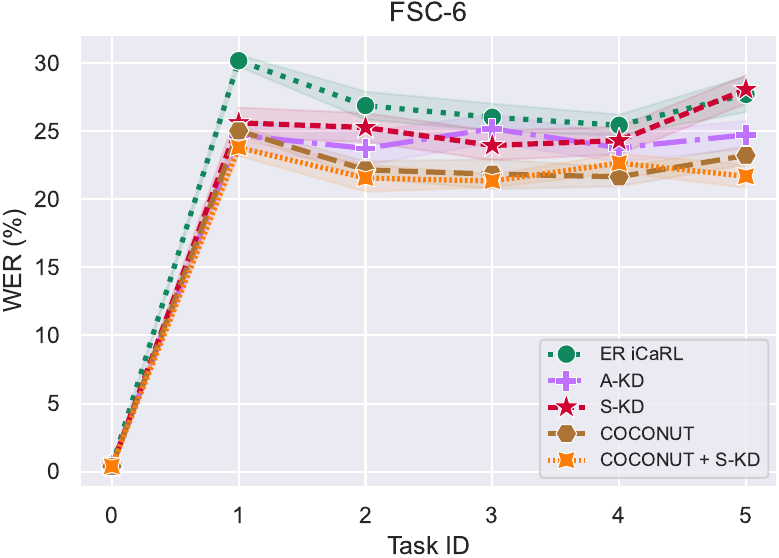}
         %\caption{$y=3\sin x$}
         %\label{fig:three sin x}
     \end{subfigure}
     \begin{subfigure}[b]{0.45\textwidth}
         \centering
         \includegraphics[width=\textwidth]{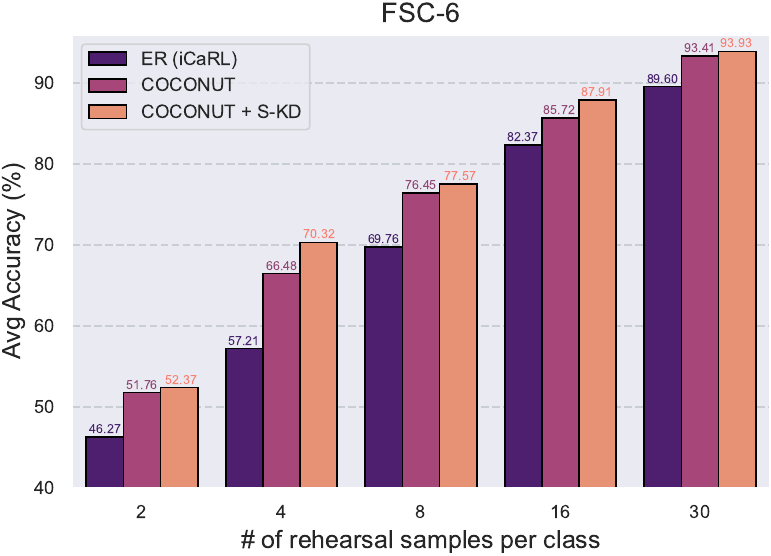}
     \end{subfigure}
    
    \caption{\textit{Left}: the trend of the WER on the observed tasks for the FSC-6 setting. \textit{Right}: the accuracy of COCONUT and other methods as a function of the memory size.}
    \label{fig:trend_taskbytask}
\end{figure*}

\textbf{Baselines}. Apart from the standard \textbf{offline} (1 task, no continual) and \textbf{fine-tuning} (no CL strategies) baselines, we compare COCONUT against standard \textbf{experience replay} (ER) methods with \textit{random} and \textit{iCaRL} \citep{rebuffi2017icarl} sampling strategies. We note that ER is already a strong baseline for FSC and SLURP. We also point out that adapting standard CL strategies to our setting is not trivial as they are usually proposed for classification tasks and not for auto-regressive tasks. Plus, we report two methods proposed in \citep{cappellazzo2023sequence} that combine rehearsal and KD principles: audio-KD (\textbf{A-KD}) that applies the KD on the audio features of the rehearsal samples, and seq-KD (\textbf{S-KD}) that, at the end of the current task, stores the text transcriptions computed with beam search only for the rehearsal samples and use them as pseudo-transcriptions for the next task. This method operates on the ASR decoder. For the sake of completeness, we also report text-KD (\textbf{T-KD}), the text counterpart of the A-KD.

\textbf{Metrics}. Following \citep{douillard2022dytox}, we report the results in terms of the \textit{Avg Acc}, which is the average of the intent accuracies after each training task, and the \textit{Last Acc}, which is the intent accuracy after the last task. We also report the \textit{Avg WER}, the average of the Word Error Rate (WER) of the extended transcription after each task. 

\subsection{Main Results}

In the first two rows of Table~\ref{tab:result_SLURPFSC}, we include the upper and lower bounds represented by the offline  learning (which is in line with the state-of-the-art) and fine-tuning approaches. For the fine-tuning approach, we can notice how CF deteriorates the knowledge of the prior classes. We then include ER baselines with buffer capacity equal to 1\% of the dataset size. From these results we can see that ER-based methods achieve good results for all metrics and configurations, confirming themselves as solid baselines. For FSC, COCONUT outperforms the other baselines by a significant margin, in terms of both accuracy and WER. Its combination with the S-KD leads to additional improvements (last row). %We also observe that COCONUT shows similar gains for both configurations (3 and 6 tasks). 

If we turn our focus to SLURP we see that, for the setting with 3 tasks, S-KD turns out to be the best approach in terms of intent accuracy, followed by COCONUT. For the WER, all the methods achieve similar performance and do not provide significant enhancements. We speculate that, as only some words are task-specific while the others are spread across multiple tasks, the text modality is less affected by CF. It is also compelling to note that the A-KD always achieves better performance than T-KD, a trend that will also be observed for the NSPT loss in the ablation studies. For SLURP-6, COCONUT slightly surpasses S-KD in terms of accuracy, and performs on par with the others for the WER metric. This indicates that COCONUT scales properly with the number of tasks. Additionally, we point out that, for SLURP, COCONUT provides less noticeable improvements than FSC. This can be attributable to the higher complexity of the dataset due to its larger dictionary and to the larger number of intents with respect to FSC (69 vs. 31). Finally, similar to FSC, the combination of COCONUT with S-KD attains the best results, confirming that fighting CF both at the encoders and ASR decoder is an effective solution.

In Fig.~\ref{fig:acc} we illustrate the trend of the intent accuracy after each task for FSC-6 and SLURP-6. For FSC-6, COCONUT outperforms the other baselines by a large margin after each task. For SLURP-6, COCONUT has a similar trend as S-KD, and their combination leads to a noteworthy boost in performance. On the left part of Fig.~\ref{fig:trend_taskbytask} we also show the trend of the WER task by task.

\subsection{Ablation Study}
\label{sec:ablation}

\textbf{Is COCONUT effective when we vary the buffer memory size?} On the right side of Fig.~\ref{fig:trend_taskbytask}, we study the trend of COCONUT for different quantities of rehearsal samples per class. Note that 8 samples per class is equivalent to a buffer capacity of 1\% of the entire training dataset. The maximum gain provided by COCONUT with respect to the ER baseline is reached for 4 and 8 samples per class ($9.27$ and $6.69$, respectively), while for the extreme cases of 2 and 30 samples, the gap is reduced. This is explained by the fact that when few samples are stored for each class, the effect of the NSPT loss is highly reduced given its reliance on the rehearsal data, whilst in the opposite case the abundance of rehearsal data makes the ER baseline already strong, thereby improving it becomes more challenging. Regarding the latter case we note that when we increase the buffer memory size, we implicitly move toward the offline setting (the upper bound), which is not the objective of this paper.
%, and this is why we decided to use 1\% as the ER ratio since we think it is a reasonable value to test our proposed approach.

\begin{table}[t]
\centering
    \caption{Ablation on the use of NSPT and NTPT losses.}

\begin{adjustbox}{width=\columnwidth,center}
\begin{tabular}{lcccc}

\toprule
\bf{Dataset} $\rightarrow$ 
 &\multicolumn{2}{c}{\CC{pastelviolet}\textbf{FSC-6}}&\multicolumn{2}{c}{\CC{champagne}\textbf{SLURP-6}} \\
 \cmidrule(r){2-3} \cmidrule(l){4-5}
\bf{Metric} $\rightarrow$ &  Avg & Last & Avg & Last\\
\bf{Method} $\downarrow$ &  Acc & Acc & Acc & Acc\\

\midrule
ER iCaRL  &69.76 &64.12 &68.08 &62.29 \\   \hdashline \addlinespace[2pt]
\rowcolor{knowcolor}
MM &71.12 &67.76 &68.78 &62.94 \\
\hdashline \addlinespace[2pt]

MM + NTPT &74.05 &67.61 &68.91 &62.57 \\ 
MM + NSPT-AA &76.30 &72.34 &69.74 &62.54\\
\rowcolor{pastelred}
MM + NSPT-AN &66.37 &63.89 &64.72 &56.84 \\
\hdashline \addlinespace[2pt]
\CC{predcolor}\textbf{MM + NSPT}&\CC{predcolor} \textbf{77.09} &\CC{predcolor} \textbf{73.80} &\CC{predcolor} \textbf{70.17} &\CC{predcolor} \textbf{63.66} \\
\bottomrule
 \end{tabular}
 \end{adjustbox}
\label{tab:ablation_MM}
\end{table}

\textbf{Ablation on the NSPT Loss}. In Table~\ref{tab:ablation_MM} we evaluate the difference in performance between the standard NTPT loss and our proposed NSPT loss and some of its variants. Specifically, we study two design properties: \textbf{1)} which samples should be used as anchors? \textbf{2)} Should the rehearsal negatives be computed using the teacher model rather than the student, unlike the negatives coming from the new task? Regarding point \textbf{(1)}, we study the case where the anchor samples are both the rehearsal data (our proposed design) \textit{and} the new data. This means that in the outer sum of Equation \ref{eq:nspt} the samples are picked from $\mathcal{I}$. Note that this design choice requires to compute the loss for all samples in the dataset, thus incurring an appreciable increase in the computational cost. We denote this variant where we \textbf{A}blate the \textbf{A}nchor design as NSPT-\textbf{AA}. As for the second point, we compute the negatives coming from the rehearsal memory using the teacher (the teacher has seen those classes in the previous tasks), whereas the samples from the current task are computed with the student model. The denominators of Equation \ref{eq:nspt} become (we use $\mathbf{z}$ to refer to both $\mathbf{a}$ and $\mathbf{t}$): $\sum_{i \in \mathcal{I}_c }\exp(\textbf{z}^{n}_k \cdot \textbf{z}^{n}_i/\tau) + \sum_{h \in \mathcal{I}_r }\exp(\textbf{z}^{n}_k \cdot \textbf{z}^{n-1}_h/\tau)$. We call it NSPT-\textbf{AN} (\textbf{A}blate \textbf{N}egatives).  

Looking at Table \ref{tab:ablation_MM}, we see that for FSC-6, the use of our proposed NSPT loss gives a considerable improvement over the NTPT loss in terms of all three considered metrics. For SLURP-6, the trend is maintained, and now the NTPT even brings a small deterioration over the MM baseline in terms of Last Acc. Also, the MM loss alone contributes positively over the ER baseline for both settings. We recall that it is not possible to study the individual contribution of the NSPT loss because, without the MM loss, the teacher projectors are randomly initialized during the second task (see section~\ref{sec:coconut}). Furthermore, we observe that the design choices of \textbf{(1)} and \textbf{(2)} are crucial to obtaining superior performance. Regarding the NSPT-\textbf{AA} loss, the model is less sensitive to this design choice. However, note that this loss is more expensive as it requires extra computational cost owing to the use of all samples in a mini-batch for its computation, thus making it less appealing than our proposed NSPT loss. Instead, the use of the NSPT-\textbf{AN} yields a severe degradation in the performance. We suspect that this happens because mixing the teacher and student at the denominators makes the learning process more complex as feature representations of different models interact, inducing more interference and thus leading the model to make more mistakes.

\begin{table}
\centering
\caption{Ablation study of the MM (upper part) and NSPT (bottom part) components. \textbf{CLS}: whether only the intent class token is used; \textbf{Anchor}: whether ER data are excluded from the anchors. \textbf{$\mathcal{L}_{\text{A}}/\mathcal{L}_{\text{T}}$}: whether the audio/text component of NSPT loss is used.}

\begin{tabular}{ccccc}
    \toprule
     \textbf{CLS} & \textbf{Anchor}  & $\mathcal{L}_{\text{A}}$ & $\mathcal{L}_{\text{T}}$ & \textbf{Avg Acc} \\
    \midrule
     &  & & & 70.10 \\ 
    \cmark & & & &70.49 \\ 
     &\cmark & & &71.09 \\ 
     \CC{knowcolor} \cmark &\CC{knowcolor} \cmark  & \CC{knowcolor} &\CC{knowcolor}&\CC{knowcolor} \textbf{71.12} \\ 
    \hdashline \addlinespace[2pt]
    \cmark & \cmark & \cmark & & 76.84 \\ 
 \cmark & \cmark & &\cmark & 73.11 \\ 
  \CC{predcolor}\cmark &\CC{predcolor} \cmark &\CC{predcolor}\cmark &\CC{predcolor}\cmark &\CC{predcolor} \textbf{77.09} \\ 
 \bottomrule
\end{tabular}%}
\label{tab:ablation_single}
\end{table}

\textbf{Ablation on the MM Loss}. Finally, in Table~\ref{tab:ablation_single} we study the design properties of the MM loss on FSC-6, and with its best configuration, we determine the individual contribution of the audio and text components to the NSPT loss. For the MM loss, we see that using the intent token and preventing the ER data from being anchors brings additional improvements. For the NSPT loss, as was evident for the A-KD and T-KD, with the former giving better results, here we also discover that the audio component is predominant. Plus, the concurrent use of both components brings a moderate increase in accuracy, and this is due to the alignment between audio and text via the MM loss.

\subsection{On the Impact of the Temperature Parameter}
\label{sec:temperature}

\begin{table}[t]
\centering
\caption{Ablation study of the temperature $\tau$ for the MM loss. We experiment on FSC-6 by setting $\tau$ beforehand and making it a learnable hyperparameter as is common practice in offline settings \citep{radford2021learning}. The light-blue row corresponds to the value we used for our experiments.}
\begin{tabular}{lccc}
\toprule
\bf{Metric} $\rightarrow$ &  Avg & Last& Avg \\
\bf{Temp. ($\tau$)} $\downarrow$ &  Acc & Acc& WER \\

\midrule
0.07 &71.06 &64.75 &\textbf{22.07}\\ \addlinespace[2pt]
\CC{predcolor}0.1  &\CC{predcolor}\textbf{71.12} &\CC{predcolor}\textbf{67.76} &\CC{predcolor}22.88 \\ \addlinespace[2pt]
0.2 &71.01 &62.35 &22.78  \\  \hdashline \addlinespace[2pt]
\CC{lightgray} Learnable &\CC{lightgray} 69.05 &\CC{lightgray} 66.33 &\CC{lightgray} 24.57 \\ 
\bottomrule
    \end{tabular}%}
\label{tab:abl_temp}
\end{table}

In this section we analyze the role of the temperature parameter in the CIL process for the MM loss (see Equation ~\ref{eq:mm}) on the FSC-6 setting. We first try to set the value beforehand ($0.07$, $0.1$, $0.2$), and then we make the temperature a learnable hyperparameter (initial value is $0.07$). Results are reported in Table~\ref{tab:abl_temp}.  We can observe that $\tau = 0.1$ is the best configuration for the accuracy metric. Note that, however, the model does not seem very sensitive to the temperature for the Avg Acc, whereas the Last Acc is more influenced. Since the Avg Acc does not change much across the three configurations, yet the Last Acc sways much more, this means that for $\tau = 0.1$ the model struggles more during the initial tasks, but it performs better towards the end of the learning process. On the other hand, learning $\tau$ task by task does not seem to be the right choice as the Avg Acc and WER metrics deteriorate with respect to the other three configurations where it is fixed. In fact, we observed that during the first tasks, the model is learning the optimal value for $\tau$ until it finds it (this value approximately lies in the range $0.134-0.142$). This initial transitional phase penalizes the accuracy of the first tasks, which in turn leads to a deterioration in the Avg Acc metric. 

\section{Conclusion}
In this work, we study the problem of E2E SLU using a seq-2-seq model for class-incremental learning. In order to mitigate catastrophic forgetting we propose COCONUT \emojismiley, a CL approach that exploits experience replay and contrastive learning paradigms. On the one hand, it preserves the previously learned feature representations via an ad-hoc supervised contrastive distillation loss, on the other it contributes to aligning audio and text representations, thus resulting in more transferable and robust to catastrophic forgetting representations. We show that COCONUT outperforms the other baselines and that synergizes with other KD techniques operating on the decoder side. We finally dissect the design choices of COCONUT through specific ablation studies, showcasing that each component is pivotal to attain the best results.

\section{Limitations}
Our work comes with some limitations. First of all, the number of suitable SLU datasets for CIL settings is limited since few datasets provide enough intent classes. Then, we could not use batches larger than $32$ owing to computational limitations, and it is known that contrastive learning  benefits from larger batches. Finally, as pointed out in the paper, almost all CIL methods are proposed for classification tasks, so their adaptation to our setting is not trivial. For this reason, we focused more on past baselines tailored for our setting, as well as rehearsal approaches that confirm themselves as strong approaches while being simple. Finally, we do not see any potential risks linked to our work.

%\section*{Acknowledgements}


\bibliography{main}

\appendix

\section{Appendix}
\label{sec:appendix}

\subsection{Related Work}

A vast array of CL strategies exist in the literature \citep{wang2023comprehensive, zhou2023deep}, which can be categorized into some macro groups: \textit{regularization}-based, \textit{experience replay}, and \textit{architecture}-based. \textit{Regularization} methods contrast forgetting either by introducing some ad-hoc regularization terms that penalize changes to model weights \citep{ebrahimi2019uncertainty, kirkpatrick2017overcoming} or to model predictions \citep{hou2018lifelong, li2017learning, fini2020online}. \textit{Experience replay} approaches interleave the new data with cherry-picked samples from the prior tasks \citep{chaudhry2018efficient,bang2021rainbow, buzzega2020dark}, or they incorporate regularization terms with this additional data to steer the optimization process and prevent catastrophic forgetting \citep{chaudhry2018efficient, wang2021training, yang2022online}. Finally, \textit{architecture} methods involve creating task-specific/adaptive parameters, such as dedicated parameters to each task \citep{xue2022meta, wang2022s} or task-adaptive sub-modules or subnetworks \citep{aljundi2017expert, ostapenko2021continual}.

Contrastive learning~\citep{oord2018representation, chen2020simple} is a popular approach in self-supervised learning, but it can also be used in supervised learning~\citep{gui2023survey} and multimodal learning~\citep{radford2021learning}.  Its objective is to learn discriminative feature representations by pushing apart different samples (negatives) and bringing closer similar ones (positives). In the case of supervised CIL, it has been shown that endowing the model with contrastive learning objectives results in more robust representations against CF. For incremental semantic segmentation, \citet{yang2022uncertainty} and \citet{zhao2023inherit} propose to exploit contrastive learning in conjunction with knowledge distillation. For image classification, \citet{wang2022online} advance a contrastive learning strategy based on the vision transformer architecture for online CL. 

\subsection{Hyper-parameters}
\label{sec:hyperparams}
We list the main hyperparameters used for our experiments in table~\ref{tab:hyperparams}. We also mention the number of epochs for each setting. For FSC-3, the number of epochs for each task is \{40,30,30\}, while for SLURP-3 we use \{40,25,25\}. For FSC-6 and SLURP-6 we use \{40,30,30,30,30,30\} and \{40,25,20,20,20,20\} epochs, respectively. We finally note that we set lr = $5\cdot 10^{-4}$ for the text encoder, the ASR decoder and the classifier, while for the audio encoder we set a smaller learning rate, lr = $5\cdot 10^{-5}$, because it is pre-trained. For our experiments, we used a single Tesla V100 or Ampere A40 GPU. Finally, each experiment reports the mean and standard deviation over $3$ runs for FSC and $2$ runs for SLURP, respectively.

\begin{table*}[hbt]
\centering
    \caption{Training hyperparameters for FSC and SLURP.}
   
    \begin{tabular}{lccc}
         \toprule
         \textbf{Hyperparameter} & \textbf{FSC} & & \textbf{SLURP} \\
         \midrule
         Batch Size & & $32$ &\\
         Optimizer & & AdamW  &\\
         $\beta_1$ & & $0.9$ & \\
         $\beta_2$ & & $0.98$& \\
         $\epsilon$ & &$10^{-6}$ & \\
         lr & & $5 \cdot 10^{-4}$ &\\
         Weight Decay & & $0.1$ &\\
         Tokenizer & Word Tok. & & BPE Tok.\\
         Beam Search width & 5 && 20 \\
         Temperature $\tau$ & & $0.1$ & \\
    \end{tabular}
    \label{tab:hyperparams}
\end{table*}

\subsection{Additional Details on the Definition of the CIL Setting for SLURP}
\label{sec:CIL}
As the SLURP dataset provides multiple levels of annotations (scenario, action, entity[es]), in principle one could decide to divide the dataset into multiple CIL tasks following one of these criteria. Following \citep{cappellazzo2023sequence}, we use the scenarios as splitting criterion because they represent more general concepts than the actions and entities, and then the accuracy is computed on the intent, defined as the pair (scenario,action). In addition to this, we define the order of the classes in the various tasks depending on their cardinality, meaning that the classes with more samples are seen first by the model. This is done because the cardinality of SLURP scenarios varies consistently from class to class, and this should resemble a practical situation in which the model accrues sufficient general knowledge, learning the largest scenarios first, that will be
useful for learning more specific scenarios. All in all, we tried to be as consistent with the original implementation in \citep{cappellazzo2023sequence} as possible in order to ensure a fair comparison with prior works.

\subsection{SpecAug Details}
\label{sec:specaug}
In this section, we elaborate on the use of SpecAug for augmenting the audio input data. SpecAug \citep{park2019specaugment} is a popular augmentation technique that is applied directly on the log mel spectrogram of an audio signal, with the aim of making the model invariant to features deformation. In the original paper, they advance three different types of distortions: \textit{time warping}, and \textit{time} and \textit{frequency masking}, where blocks of consecutive time steps and frequency channels are zero-masked, respectively. Since our audio encoders (i.e., DistilHuBERT and Wav2vec 2.0) work on the raw audio waveforms, SpecAug is not applicable by default. In order to circumvent this problem, we apply an approximated version of SpecAug directly to the raw waveform, as proposed in the SpeechBrain library \citep{ravanelli2021speechbrain}. We randomly drop chunks of the audio waveform (by zero-masking) and frequency bands (with band-drop filters). Unlike the SpeechBrain implementation, we do not apply speed perturbation. In more detail, with probability $0.5$ we randomly drop up to $2$ frequencies, while with probability $0.5$ we randomly drop up to $3$ chunks of audio whose length is sampled from a uniform distribution $\sim \mathcal{U}(0,0.05 \cdot len(x))$, where $len(x)$ is the length of the considered audio waveform $x$.

\begin{figure*}
     \centering
     \begin{subfigure}[b]{0.45\textwidth}
         \centering
         \includegraphics[width=\textwidth]{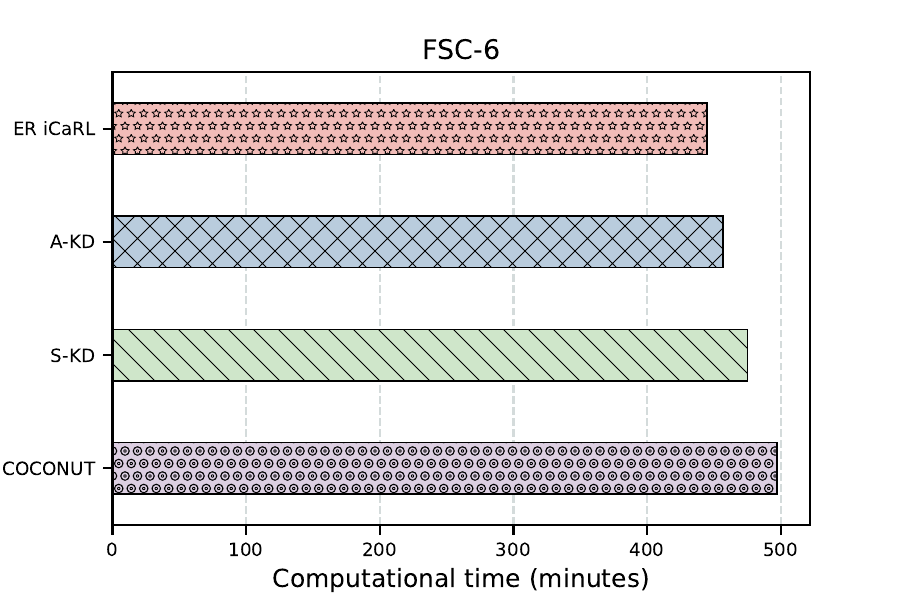}
         %\caption{$y=3\sin x$}
         %\label{fig:three sin x}
     \end{subfigure}
     \begin{subfigure}[b]{0.45\textwidth}
         \centering
         \includegraphics[width=\textwidth]{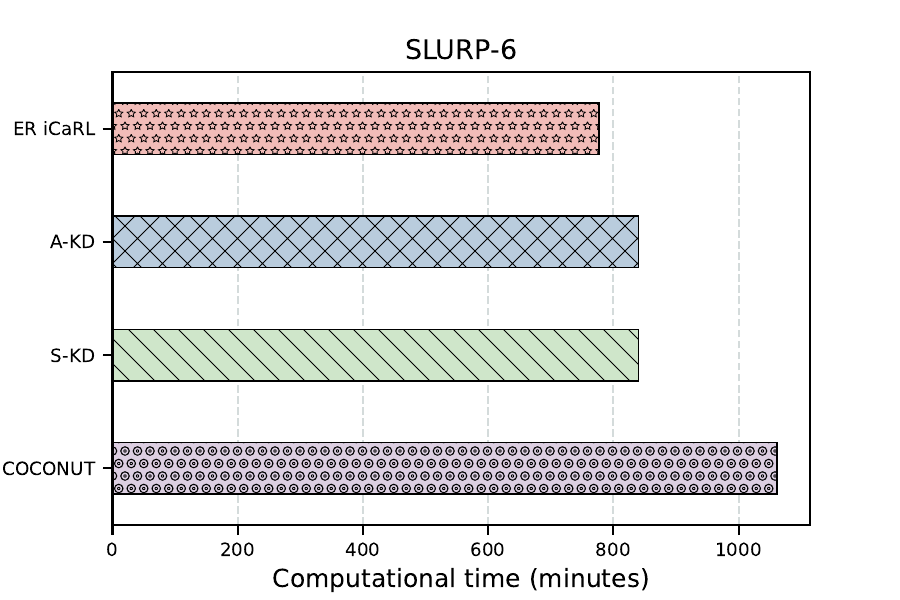}
     \end{subfigure}
    
    \caption{Computational cost analysis of various CIL methods for FSC-6 (\textit{left}) and SLURP-6 (\textit{right}).}
    \label{fig:compute_time}
\end{figure*} 

\subsection{Additional Results for SLURP}

Similar to the study we proposed on the right side of Fig.~\ref{fig:trend_taskbytask} for the FSC dataset, we here include a similar one for SLURP, where we vary the number of samples per class stored in the rehearsal memory. We report these additional results in Fig.~\ref{fig:varmemory_slurp}. Note that $1260$ samples corresponds to 1\% of the training data, which is the \% we used for our main results. Similar to what we obtained for the FSC dataset, we see that, as we increase the number of retained samples to 2500, the gain brought by COCONUT and its combination with S-KD is a bit smaller but still significant, and this happens because the iCaRL method becomes a stronger and stronger baseline as we increase the \% of data. Also, we notice that adding the S-KD approach is more beneficial when we have fewer samples in the memory since the task is way more challenging.

\begin{table}
\centering
\caption{The accuracy of COCONUT and other methods as a function of the memory size for the SLURP dataset.}

\begin{tabular}{lccc}
    \toprule
     & \multicolumn{3}{c}{\textbf{Examples per class}} \\
     \cmidrule(r){2-4}
     \textbf{Method} & \cellcolor{almondlow}650  & \cellcolor{almondmiddle} 1260 & \cellcolor{almondultra}2500 \\
    \midrule
    iCaRL & 59.94 & 61.87 & 63.38 \\
    COCONUT & 68.08 & 70.17 & 71.91 \\
   \hdashline  \addlinespace[2pt]
    COCONUT + S-KD & \cellcolor{knowcolor} \textbf{70.15} & \cellcolor{knowcolor} \textbf{71.41} & \cellcolor{knowcolor} \textbf{72.10} \\
 \bottomrule
\end{tabular}%}
\label{fig:varmemory_slurp}
\end{table}

\subsection{Computational Time Analysis}
In this section, we study the computational cost of COCONUT and compare it with the other baselines. The computational time includes the training and inference time, as well as the time needed for selecting the rehearsal samples to store in the memory (the S-KD method also computes the pseudo-labels that will be stored in the memory). The main difference between the baselines (ER iCaRL, A-KD, S-KD) and COCONUT is that the baselines focus on the rehearsal data only, while COCONUT is applied to both the rehearsal data (NSPT loss) and the new data (MM loss), and so COCONUT requires an additional compute time due to the MM loss. Nevertheless, this additional time does not hinder its applicability as it is somewhat limited. Indeed, for the FSC-6 setting, the KD baselines require an additional 3/7 \% of computational time with respect to the fine-tuning baseline, while COCONUT requires around 11\%. For SLURP-3, the KD baselines require around 8\% of additional compute time, whereas COCONUT requires around 35\%. Undoubtedly COCONUT requires slightly more running time than the other KD baselines that are only applied to the rehearsal samples, but this overhead is minimal and consequently we believe this is not an issue for a practical scenario, considering also that COCONUT leads to much-improved performance. Additionally, from a memory overhead point of view, COCONUT requires the storage of the rehearsal samples and a copy of the model from the previous task. These storage requirements are the same as the A-KD baseline. Instead, the S-KD approach, in addition to the aforementioned storage requirements, also necessitates the storage of the rehearsal text transcriptions generated with beam search from the previous task, thus increasing the requested memory overhead with respect to COCONUT.

\subsection{Future Work}
COCONUT relies on two contrastive learning-based losses applied to the projections of audio and text encoders outputs. In principle, COCONUT could be exploited in other multi-modal settings such as audio-vision or vision-language. Therefore, it would be interesting to study whether COCONUT can be exploited in other different multi-modal scenarios. Also, since these settings usually involve a larger number of classes than ours, we would be able to test how COCONUT scales to the number of tasks.

\end{document}